\documentclass[3p,times,procedia]{elsarticle}
\usepackage{nupha_ecrc}

\volume{00}

\firstpage{1}

\journalname{Nuclear Physics A}

\runauth{}

\jid{nupha}

\jnltitlelogo{Nuclear Physics A}

\usepackage{amssymb}

\usepackage[figuresright]{rotating}

\newcommand {\Ncoll}	{N_{\rm coll}}
\newcommand {\mean}[1]	{\langle #1\rangle}

\begin{document}

\begin{frontmatter}


\dochead{}
%% e.g. \dochead{17th International Conference on Dynamical Processes in Excited States of Solids}

\title{Elliptic Anisotropy $v_2$ May Be Dominated by Particle Escape
  instead of Hydrodynamic Flow}

\author{Zi-Wei Lin,$^a$ Liang He,$^b$ Terrence Edmonds,$^c$ Feng Liu,$^d$ Denes Molnar,$^b$ Fuqiang Wang $^{b,d}$}
\address{$^a$ Department of Physics, East Carolina University, Greenville, NC 27858, USA}
\address{$^b$ Department of Physics and Astronomy, Purdue University, West Lafayette, IN 47907, USA}
\address{$^c$ Department of Physics, University of Florida, Gainesville, FL 32611, USA}
\address{$^d$ Key Laboratory of Quark \& Lepton Physics (Central China Normal University), Ministry of Education, China}

\begin{abstract}
It is commonly believed that azimuthal anisotropies in relativistic
heavy ion collisions are generated by hydrodynamic evolution of the
strongly interacting quark-gluon plasma. 
Here we use transport models to study how azimuthal anisotropies
depend on the number of collisions that each parton suffers. 
We find that the majority of $v_2$ comes from the
anisotropic escape of partons, not from the parton collective flow,
for semi-central Au+Au collisions at 200A GeV. As expected, the
fraction of $v_2$ from the anisotropic particle escape is even higher
for smaller systems such as d+Au. Our transport model results also
confirm that azimuthal anisotropies would be dominated by hydrodynamic
flow at unrealistically-high parton cross sections. 
Our finding thus naturally explains the similarity of azimuthal
anisotropies in small and large systems; however, it presents a
challenge to the paradigm of anisotropic flow. 

\end{abstract}


\end{frontmatter}

\section{Introduction}
\label{Sec-intro}

A main goal of relativistic heavy ion collisions is to create the
quark-gluon plasma (QGP) and then study its properties in order to
better understand quantum chromodynamics at extreme 
conditions. In non-central heavy ion collisions, the overlap volume of the
colliding nuclei is anisotropic in the transverse plane 
even when event-by-event fluctuations in the initial geometry are
neglected, and large final-state azimuthal anisotropies in the
momentum space have been measured and described 
well by hydrodynamics \cite{Heinz:2013th}. Transport models such as
A Multi-Phase Transport (AMPT) \cite{Lin:2004en} can also
describe the observed final-state anisotropic flow. 
It is thus commonly believed that azimuthal anisotropies (in large
colliding systems at least) are generated by the hydrodynamic evolution of
the strongly-interacting QGP and that the evolution of a large system
in transport models is qualitatively the same as that in hydrodynamics.

Recently, anisotropy signals have been observed in small
systems of p+Pb collisions at the LHC \cite{Khachatryan:2010gv} 
and d+Au collisions at RHIC \cite{Adare:2014keg}, and the signals 
can again be described by both hydrodynamics \cite{Bozek:2010pb} and
the AMPT transport model \cite{Bzdak:2014dia}.
A natural question is whether hydrodynamics is still applicable to
such small systems. So we investigate the generation of azimuthal
anisotropies in transport models by analyzing all parton collisions in 
semi-central Au+Au collisions and central d+Au collisions at
$\sqrt s=200A$ GeV \cite{He:2015hfa}. 
We use the string melting version of the AMPT model
\cite{Lin:2001zk,Lin:2004en}, and parton scatterings are treated with
the ZPC elastic parton cascade \cite{Zhang:1997ej}. We analyze the
complete history of parton interactions in AMPT from the initial
encounter of the colliding nuclei to the final-state partons.
We also use another transport model, the MPC/Cascade
\cite{Molnar:2001ux}, to check the model dependence of our results. 

\section{Results}
\label{Sec-escape}

\begin{figure}[htb]
\begin{minipage}[t]{70mm}
\includegraphics[height=2.05in]{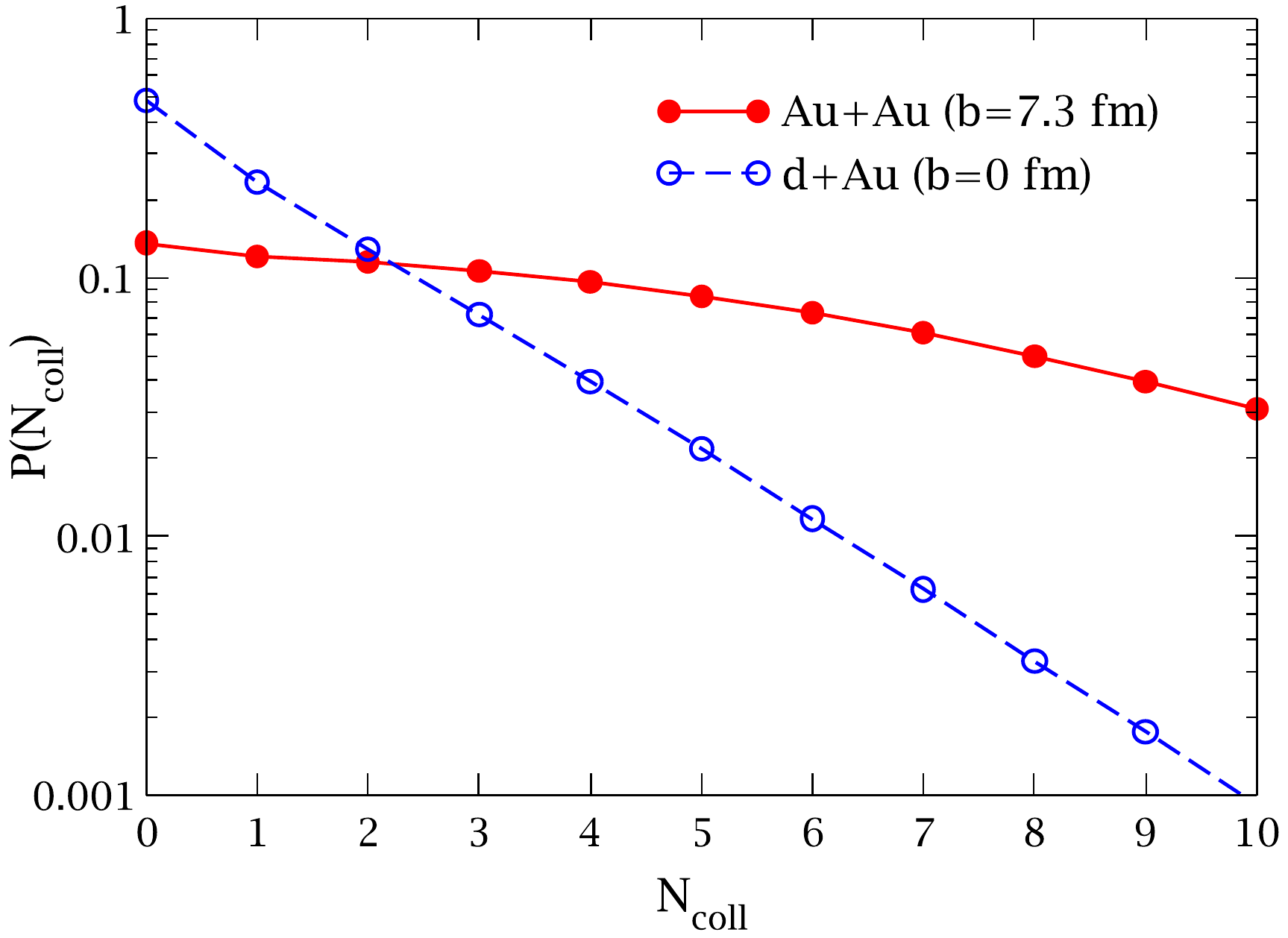}
\vspace{-0.6cm}
\caption{(color online) Normalized probability distributions of
  partons that freeze out after $\Ncoll$ collisions: solid and dashed
  curves are for Au+Au and d+Au collisions, respectively.} 
\label{fig1}
\end{minipage}
\hspace{\fill}
\begin{minipage}[t]{70mm}
\includegraphics[height=2.03in]{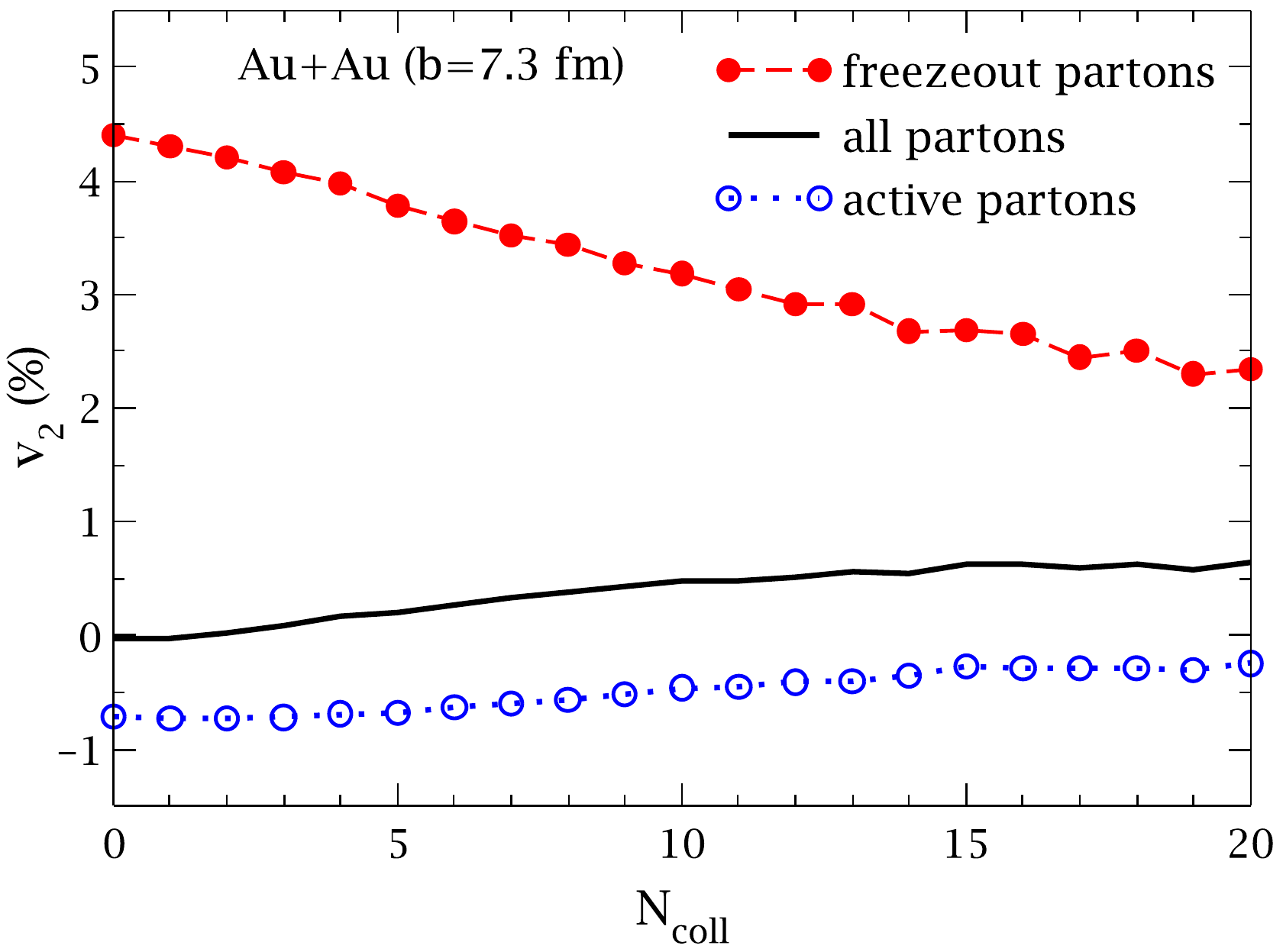}
\vspace{-0.6cm}
\caption{Parton $v_2$ in Au+Au collisions versus $\Ncoll$: 
dashed curve for partons that freeze out after exactly $\Ncoll$
collisions, dotted curve for partons continuing to interact
after $\Ncoll$ collisions, solid curve for the sum of these two
parton populations.
} 
\label{fig2}
\end{minipage}
\end{figure}

We define $\Ncoll$ as the number of collisions that a parton 
suffers with other partons. Fig.\ref{fig1} shows the normalized
probability distributions of partons that freeze out after $\Ncoll$
collisions for semi-central Au+Au collisions at the impact parameter 
$b=7.3$ fm (solid curve)  and central d+Au collisions at $b=0$ fm (dashed
curve). We see that about 14\% of partons in the Au+Au collisions
do not scatter at all before they freeze out, while this fraction is
much higher at 48\% for d+Au collisions. 
Summing over all possible $\Ncoll$ values, we get the average
number of collisions as $\mean{\Ncoll}$=4.6 for Au+Au collisions 
($b=7.3$ fm) and 1.2 for d+Au collisions ($b=0$ fm) respectively.

Fig.\ref{fig2} shows the parton $v_2$ as a function of the number of
collisions $\Ncoll$ for different parton populations in Au+Au
collisions. The dashed curve shows $v_2$ of partons that freeze out after exactly
$\Ncoll$ collisions, while the dotted curve is for partons that
continue to scatter after $\Ncoll$ collisions. The sum of these two
parton populations is shown by the solid curve and labeled ``all
partons''; however, note that it represents all partons that have
already scattered $\Ncoll$ times, not all partons in the
collisions (except when $\Ncoll=0$). 
Fig.\ref{fig2} shows that freezeout partons at $\Ncoll=0$ have a large
positive $v_2$; this is fully due to the anisotropic escape probability 
since it is easier for a parton to escape (i.e. freeze out) along the
shorter axis of the transverse overlap area. The solid curve at
$\Ncoll=0$ shows zero $v_2$ because the initial all-parton $v_2$ is zero due
to the azimuthal symmetry in the transverse momentum
distribution of initial partons. 
It is interesting to see that active partons at $\Ncoll=0$, i.e., all
partons that will scatter once or more times during the parton
evolution, have an initial $v_2$ that is negative. However, this is
only natural since at $\Ncoll=0$ the $v_2$ of active partons must have the
opposite sign as the $v_2$ of freezeout partons
so that the total initial $v_2$ is zero.
We also see that the same qualitative features at $\Ncoll=0$
are repeated at higher $\Ncoll$ values in Fig.\ref{fig2}, i.e.,
partons that freeze out after a given $\Ncoll$ collisions continue to
have a significant positive $v_2$, active partons at a given $\Ncoll$
continue to have a negative $v_2$, while the sum has a small or near-zero $v_2$. 

\begin{figure}[htb]
\includegraphics[height=2.3in]{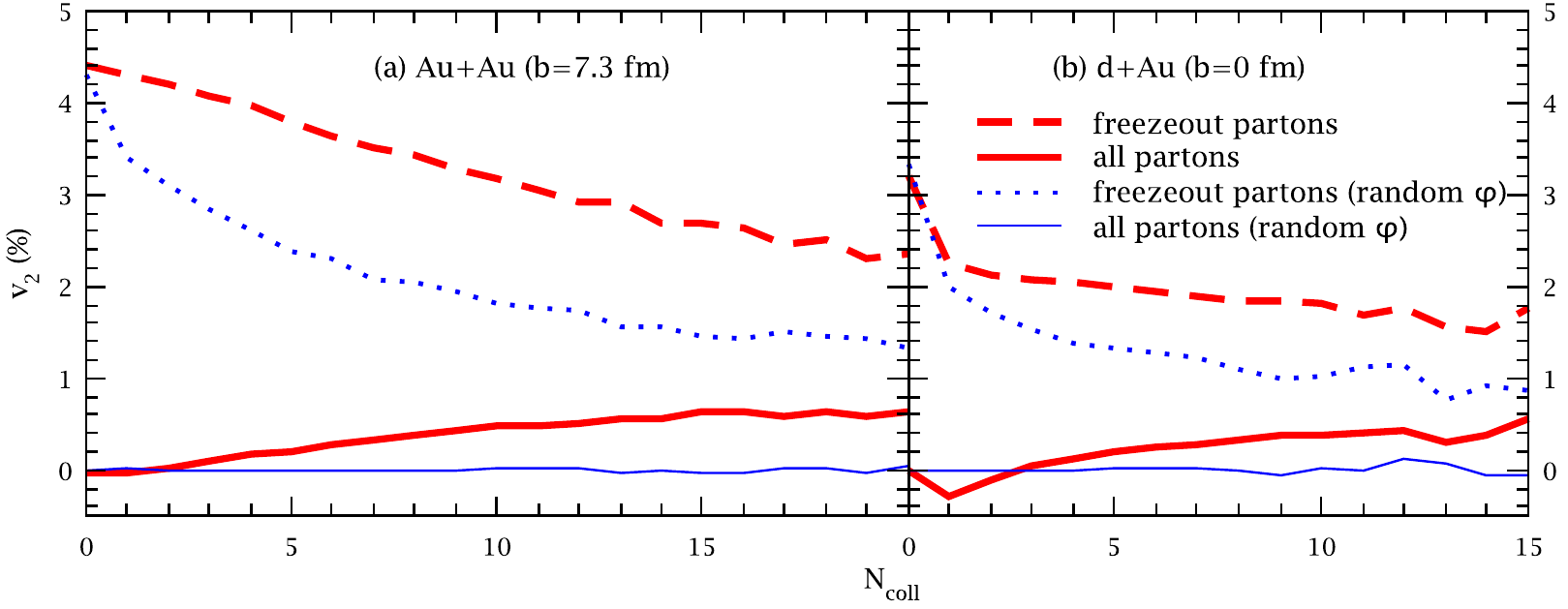}
\vspace{-0.8cm}
\caption{Parton $v_2$ as a function of the number of collisions
  $\Ncoll$ in (a) Au+Au and (b) d+Au collisions  from both normal
  (thick curves) and azimuth-randomized (thin curves) AMPT results.
} 
\label{fig3}
\end{figure}

The final average elliptic flow of all partons $\mean {v_2}$
includes partons that freeze out at different $\Ncoll$ ($0, 1,
2, ...$), but freezeout partons at $\Ncoll \neq 0$ obtain a positive
$v_2$ due to two effects: 1) the anisotropic escape probability due to
spatial anisotropy, and 2) their anisotropic collective flow (or the
space-momentum correlation) that is generated by parton
scatterings during the evolution. 
So a key question is which effect is more important. Note
that both effects are due to parton scatterings and thus inseparable
in the space-time evolution. In order to estimate
the contribution from the escape mechanism, i.e., from
the anisotropic escape probability if there were no space-time
correlation in the partons during the evolution, 
we design the azimuth-randomized test \cite{He:2015hfa}. 
In test simulations, we randomize the $\phi$ angle while keeping
the same magnitude for the transverse momentum of each of the two
final-state partons after each parton scattering.

As shown by the thin solid line in Fig.\ref{fig3}(a) for Au+Au
collisions and in Fig.\ref{fig3}(b) for d+Au collisions, 
the ``all partons'' $v_2$ in the azimuth-randomized test 
is essentially zero at any given $\Ncoll$ value. 
This confirms that we have destroyed the collective flow of partons
that have suffered any given number of scatterings $\Ncoll$. 
Some of these partons will not scatter again and thus freeze out after
exactly $\Ncoll$ scatterings; they still have a significant positive
$v_2$, as shown by the dotted lines in Fig.\ref{fig3}. Since the random test
removes the azimuthal preference and space-momentum correlation due to
the collective flow, the freezeout parton $v_2$ comes purely from the
escape mechanism. Fig.\ref{fig3} also shows that the freezeout parton
$v_2$ in the random test is lower than the normal AMPT results (dashed
curves) at $\Ncoll \neq 0$; this is expected since the random test 
results lack the $v_2$ contribution from the collective flow. On the
other hand, at $\Ncoll=0$ the freezeout parton $v_2$ 
in the random test is essentially the same as the normal results; this
is because it is determined only by the initial spatial anisotropy,
which is not affected by the azimuth randomization. 

Summing the freezeout parton $v_2$ over all possible $\Ncoll$ values
with freezeout probabilities (shown in Fig.\ref{fig1} for normal events
and similar for the random tests) as the weight, we get the final
average elliptic flow $\mean {v_2}$ at 3.9\% for normal events and 
2.7\% for azimuth-randomized events for the Au+Au collisions. 
Fig.\ref{fig3}(b) shows AMPT results for central d+Au collisions, 
which are qualitatively the same as those for Au+Au collisions,
while $\mean {v_2}$ is 2.7\% for normal events and 2.5\% for
azimuth-randomized events. Therefore a majority of $v_2$ comes from
the escape mechanism for both collision systems. 
In addition, although the $v_2$ magnitudes for freezeout partons are lower in d+Au
collisions as expected, they are not drastically lower despite a
factor of $\sim 4$ lower $\mean{\Ncoll}$. This is because partons obtain
their $v_2$ mainly through successive freezeout (at different $\Ncoll$) 
in response to the spatial anisotropy, not mainly from the
hydrodynamic-type flow that gradually builds up after many scatterings.

\begin{figure}[htb]
\begin{minipage}[t]{70mm}
\includegraphics[width=2.7in]{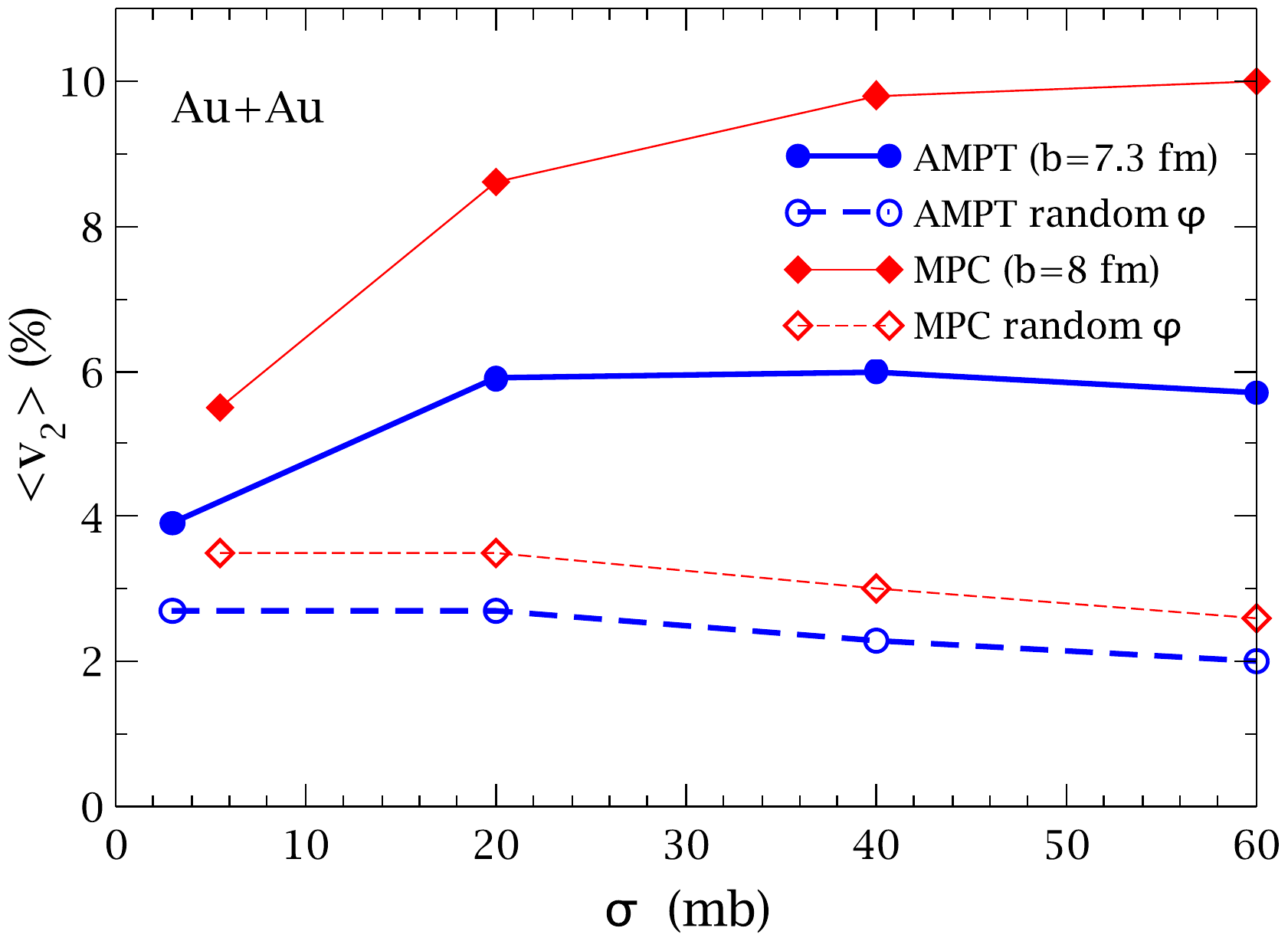}
\vspace{-0.3cm}
\caption{Parton $v_2$ as a function of the parton cross section; 3 mb
  is the value at which the string melting AMPT model 
reasonably reproduces bulk data including pion and kaon $v_2(p_T)$.}
\label{fig4}
\end{minipage}
\hspace{\fill}
\begin{minipage}[t]{70mm}
\includegraphics[width=2.75in]{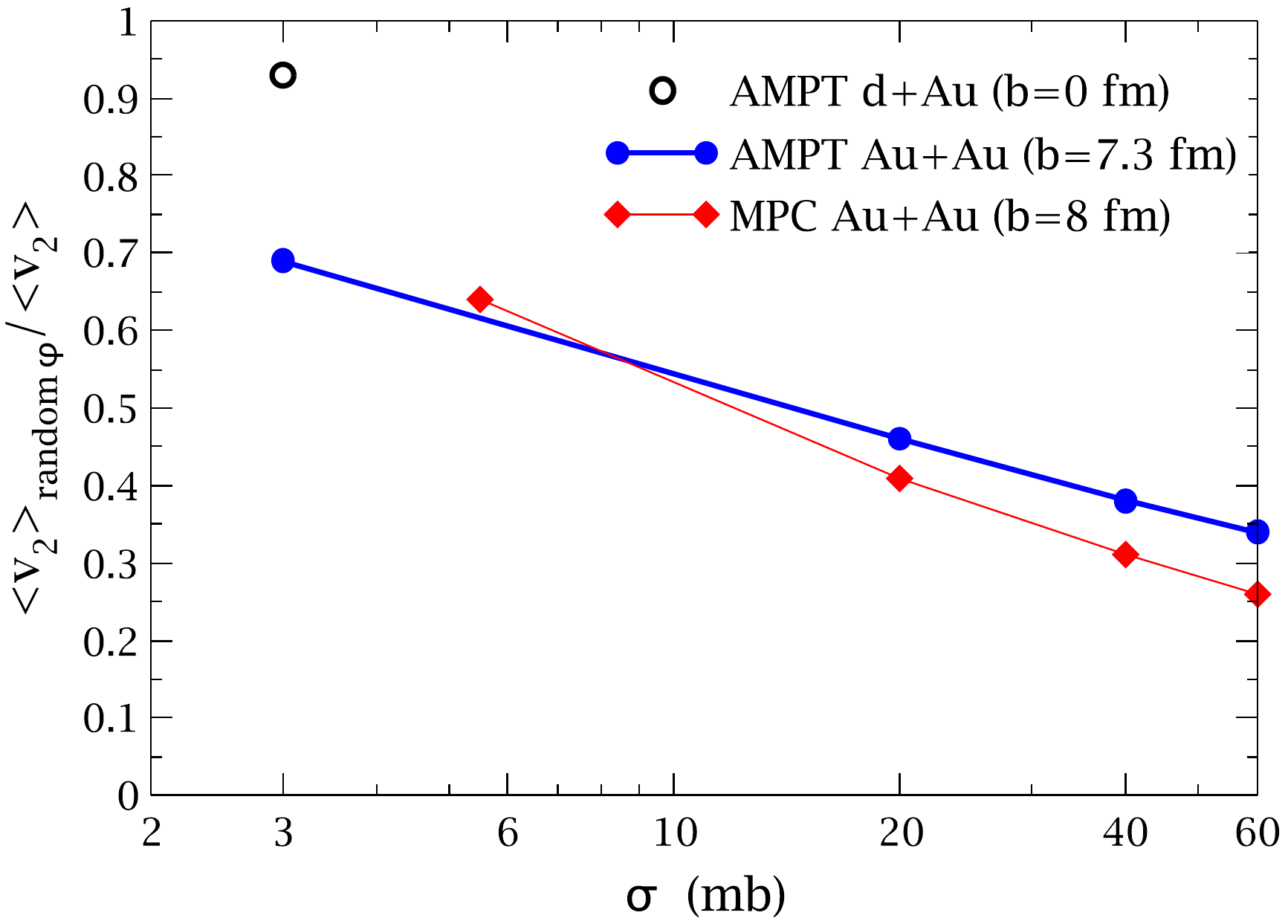}
\vspace{-0.7cm}
\caption{Parton $v_2$ ratios as functions of the parton
  cross section for d+Au and Au+Au collisions.} 
\label{fig5}
\end{minipage}
\end{figure}

The parton cross section used in the AMPT results shown above is
$\sigma=3$ mb, which is  determined from comparisons with the bulk
pion and kaon data on dN/dy, $p_T$ spectra and $v_2(p_T)$ in heavy ion
collisions at $200A$ GeV and $2760A$ GeV \cite{Lin:2014tya}. 
We have seen that the corresponding average number of collisions
$\mean{\Ncoll}$ is modest (4.6) for semi-central Au+Au collisions and
low (1.2) for central d+Au collisions. 
On the other hand, at high-enough $\mean{\Ncoll}$ we expect transport
models to approach hydrodynamics, where we can also expect the
final $\mean{v_2}$ to be dominated by hydrodynamic-type flow instead
of the escape mechanism. Therefore we have changed $\sigma$ in
the AMPT model arbitrarily up to 60 mb \cite{He:2015hfa} with the results shown
in Fig.\ref{fig4} (curves with circles). MPC results are also shown
(curves with diamonds), where $\mean{v_2}$ values are bigger partly
due to the usage of an isotropic $d\sigma/dt$. Note that $d\sigma/dt$ in
AMPT has a Debye-screened form \cite{Lin:2004en}. We see in Fig.\ref{fig4} that
$\mean{v_2}$ from the normal transport results increases or roughly
saturates when the cross section increases, while $\mean{v_2}$ from
the azimuth-randomized tests decreases with the cross section. 

Fig.\ref{fig5} shows the ratio of the azimuth-randomized $\mean{v_2}$
over the normal $\mean{v_2}$, which can be considered as an estimate
of the fraction of $\mean{v_2}$ from the escape mechanism. As expected,
this ratio from either AMPT or MPC decreases with $\sigma$ (roughly 
linear with $\ln\sigma$), verifying our expectation that $\mean{v_2}$
is eventually dominated by hydrodynamic-type flow at very high cross
sections. However, at $\sigma=3$ mb 
which enables the AMPT model to reasonably reproduce the bulk data 
\cite{Lin:2014tya}, $\mean{v_2}$ is dominated by the escape mechanism.

%\section{Discussions and Summary}
\section{Summary}
\label{Sec-summary}

Using two different transport models, we find that the majority of
$v_2$ comes from the anisotropic escape of partons, not from the
parton collective flow, even for semi-central Au+Au collisions at
$200A$ GeV. This is more true for smaller systems such as central d+Au
collisions. The domination of $v_2$ by the escape mechanism 
corresponds to low and modest $\mean{\Ncoll}$, the average number of
scatterings each parton suffers. At very large $\mean{\Ncoll}$ (or at 
unrealistically large parton cross sections), on the other hand, we find
that hydrodynamic-type collective flow will eventually dominate. 
So the key question is which case heavy ion collisions in RHIC and LHC
energies belong to, and further studies are needed in order to
identify unique signatures of the escape mechanism. 
If confirmed, our finding could change the paradigm of anisotropic flow. 
In addition, we find that
the escape mechanism still contributes $\sim 30\%$ of the final parton
$\mean{v_2}$ even at the parton cross section of 60 mb. 
Since parton escape is inevitable for transient collision
systems, it is also important to examine the possible role of the
escape mechanism in the hydrodynamics framework. 

%% References

\bibliographystyle{elsarticle-num}
%\bibliography{mybib}

\end{document}